\documentclass[conference]{IEEEtran}
\IEEEoverridecommandlockouts
\usepackage{cite}
\usepackage{amsmath,amssymb,amsfonts}
\usepackage{algorithmic}
\usepackage{graphicx}
\usepackage{mathptmx}
\usepackage{textcomp}
\usepackage{xcolor}
\usepackage{url}
\usepackage{graphicx}
\usepackage{float}
\usepackage[justification=centering]{caption}
\usepackage{booktabs}
\usepackage{booktabs}
\usepackage{booktabs}
\usepackage{graphicx}
\usepackage{subcaption}

\def\BibTeX{{\rm B\kern-.05em{\sc i\kern-.025em b}\kern-.08em
    T\kern-.1667em\lower.7ex\hbox{E}\kern-.125emX}}

\begin{document}
\IEEEoverridecommandlockouts\IEEEpubid{\makebox[\columnwidth]{ 979-8-3503-5171-2/24/\$31.00 $\copyright$2024 IEEE \hfill}\hspace{\columnsep}\makebox[\columnwidth]{ }}
\title{6Vision: Image-encoding-based IPv6 Target Generation in Few-seed Scenarios\\}
\DeclareRobustCommand*{\IEEEauthorrefmark}[1]{%
    \raisebox{0pt}[0pt][0pt]{\textsuperscript{\footnotesize\ensuremath{#1}}}}
\author{
	\IEEEauthorblockN{
		Wenjian Zhang\IEEEauthorrefmark{1}, 
		Guanglei Song\IEEEauthorrefmark{2}, 
		Lin He\IEEEauthorrefmark{1,2}, 
		Jinlei Lin\IEEEauthorrefmark{1}, 
		Songyun Wu\IEEEauthorrefmark{1},\\
            Zhiliang Wang\IEEEauthorrefmark{1,2},
            Chenglong Li\IEEEauthorrefmark{1,2},
            Jiahai Yang\IEEEauthorrefmark{3,1,2}
            }
	\IEEEauthorblockA{\IEEEauthorrefmark{1}Institute for Network Sciences and Cyberspace, BNRist, Tsinghua University, Beijing, China}
	\IEEEauthorblockA{\IEEEauthorrefmark{2}Zhongguancun Laboratory, Beijing, China}
        \IEEEauthorblockA{\IEEEauthorrefmark{3}Quan Cheng Laboratory, Shandong, China}

} 

\maketitle

\begin{abstract}
Efficient global Internet scanning is crucial for network measurement and security analysis. While existing target generation algorithms verify remarkable performance in large-scale detection, their efficiency notably diminishes in few-seed scenarios. This decline is primarily attributed to the intricate
configuration rules and sampling bias of seed addresses.
Moreover, instances where BGP prefixes have few seed addresses are widespread, constituting 63.65\% of occurrences.
We introduce 6Vision to tackle this challenge by introducing a novel approach to encoding IPv6 addresses into images, facilitating comprehensive analysis of intricate configuration rules. Through feature stitching, 6Vision not only improves the learnable features but also amalgamates addresses associated with configuration patterns for enhanced learning. Moreover, it integrates an environmental feedback mechanism to refine model parameters based on identified active addresses, thereby alleviating the sampling bias inherent in seed addresses. As a result, 6Vision achieves high-accuracy detection even in few-seed scenarios.
The \textit{HitRate} of 6Vision is improved by 181\%$\sim$2,490\% compared to existing algorithms, while the \textit{CoverNum} is 1.18$\sim$11.20 times that of them. Additionally, 6Vision can function as a preliminary detection module for existing algorithms, yielding a conversion gain (\textit{CG}) ranging from 242\%$\sim$2,081\%. Ultimately, we achieve a conversion rate (\textit{CR}) of 28.97\% for few-seed scenarios. We enrich the IPv6 hitlist, not only enhancing current target generation algorithms for large-scale address detection in few-seed scenarios but also effectively supporting IPv6 network measurement and security analysis.
\end{abstract}


\section{Introduction}\label{chap:introduction}
Active address scanning technology is a fundamental prerequisite for large-scale network measurement. 
For example, the probed active addresses can be used to measure network topology to reflect the interconnections of nodes\cite{beverly2018ip,vermeulen2020diamond}, probe Internet services for comprehensive assessments and resource\cite{durumeric2013zmap,izhikevich2021lzr,fan2022autoiot,pan2023your,song2023doors}, census\cite{heidemann2008census,fukuda2018knocks}, and assess network security by evaluating the attack surface\cite{li2021fast,saidi2022one,moon2021accurately,liu2020addressless}. 

Due to its rapid scanning capabilities, active address scanning can be swiftly conducted within the IPv4 space. For example, Masscan \cite{masscan} can scan active addresses on specific ports across the IPv4 network in just 5 minutes.
As the Internet rapidly evolves into a global critical infrastructure, IPv4 is no longer sufficient to meet its expansion requirements, resulting in the accelerated deployment of IPv6 worldwide. For instance, as of May 3, 2024, over 43.21\% of users accessed Google \cite{Goggle} through IPv6. However, the vast address space of IPv6 presents significant challenges for active address scanning. Utilizing currently fast scanners ZMap\cite{Zmapv6} and Masscan\cite{masscan} to brute-force scan the entire IPv6 address space would necessitate at least several million years\cite{borgolte2018enumerating}. Therefore, brute-force scanning of the entire IPv6 address space is impractical.

In response to the impracticality of brute-force scanning the entire IPv6 address space, researchers have developed efficient target generation algorithms\cite{ullrich2015reconnaissance,foremski2016entropy,murdock2017target,liu20196tree,hou20216hit,hou20236scan,gasser2018clusters,song2022det,yang20226graph,yang20226forest,song2022addrminer,hou2023search}. These algorithms are designed to learn the configuration patterns of seed addresses (i.e., known addresses) and generate target addresses with a high probability of being active.
Existing target generation algorithms have attained high detection accuracy in scenarios where an ample number of seed addresses are available. 
However, we find that their accuracy is significantly lower than claimed in few-seed scenarios. For example, the target generation algorithm DET \cite{song2022det} achieves an accuracy rate of over 25\% in scenarios with sufficient seed addresses, but only 3.73\% in few-seed scenarios. In scenarios with sufficient seed addresses, the abundance of seed addresses enables the target generation algorithm to approximate the address configuration of the real network more closely, thereby diminishing the severity of sampling bias
\cite{ullrich2015reconnaissance,foremski2016entropy,murdock2017target,liu20196tree,hou20216hit,hou20236scan,gasser2018clusters,song2022det,yang20226graph,yang20226forest,song2022addrminer}. 
What's more, existing target generation algorithms tend to probe within a limited number of patterns in scenarios with sufficient seed addresses, as these few patterns encompass a large number of addresses. 
However, due to the \textbf{sampling bias} and \textbf{intricate configuration rules} associated with seed addresses in few-seed scenarios, existing target generation algorithms struggle to accurately mine and often misinterpret the configuration rules. 
Moreover, we have found that 63.65\% of BGP prefixes have only a few seed addresses, indicating that this is a common issue. For more details, see \S \ref{chap:few-seed}.

Addressing the two aforementioned issues presents a highly challenging task. Firstly, the configuration pattern of seed addresses is exceedingly intricate. In few-seed scenarios, only a handful of seed addresses exist under each BGP prefix, and there is substantial variation in the configuration of seed addresses across different BGP prefixes. Secondly, the extremely limited number of seed addresses results in severe sampling bias. Consequently, target generation algorithms relying solely on pattern extraction are too simplistic to operate effectively.
Our central approach to addressing these challenges is to meticulously unveil the concealed associations within an address at a finer granularity and to mine the associations between addresses. This enables effective detection even in few-seed scenarios. Additionally, we dynamically adjust the detection direction based on feedback from detected active addresses, thereby mitigating sampling bias.

6Vision introduces a novel approach by encoding addresses into images for the first time. Each address is encoded into an 8*16 image, and the IPv6 addresses are rearranged by groups
\footnote{The four hexadecimal characters are a group, separated by a colon in an IPv6 address.}
, with each row representing a group of IPv6 addresses. Additionally, each column corresponds to the same position in each group, facilitating a comprehensive representation of the address structure. This encoding method amalgamates similar positions within the address, thereby enabling a more fine-grained exploration of the configuration patterns present in the address. Previous studies have demonstrated that learning similar features can enhance learning outcomes \cite{cui20216gan}. Therefore, 6Vision employs clustering to group seed addresses with similar image features into subclasses after image encoding. This approach enables subsequent learning modules to efficiently learn the image features of addresses within each subclass. Next, 6Vision employs feature stitching to augment the number of learnable image features. It combines image encodings configured with similar addresses, thereby enabling subsequent modules to learn the relationships between addresses. This approach allows the module to uncover the intricate configuration rules of seed addresses in few-seed scenarios. Subsequently, 6Vision uses autoregressive-based Gated PixelCNN \cite{van2016conditional} to learn and generate image features, then decode the generated image features into target addresses.
Finally, 6Vision introduces an environment feedback mechanism to mitigate the impact of sampling bias. This mechanism enables fine-tuning of model parameters based on detected active addresses, thus enhancing the accuracy of the model. As the investigation progresses, the model learns an increasing number of addresses. Consequently, the patterns of seed addresses learned by the model tend to converge with the configuration patterns present in the real network.
6Vision demonstrates high-accuracy detection capabilities even in few-seed scenarios. Additionally, it can serve as a preliminary detection module for existing target generation algorithms. This module expands the number of addresses and mitigates the sampling bias inherent in seed addresses, thereby facilitating the exploration of patterns.

The contributions of this paper are as follows:
\begin{itemize}
    \item We integrate computer vision techniques with IPv6 active address detection through 6Vision. By leveraging feature stitching and environment feedback mechanisms, 6Vision successfully addresses challenges encountered in few-seed scenarios and achieves high-accuracy detection. Furthermore, we demonstrate the capability to represent the configuration of an address or an address set using an image.
    \item  The \textit{HitRate} of 6Vision is improved by 181\%$\sim$2,490\% compared to existing algorithms, while the \textit{CoverNum} is 1.18$\sim$11.20 times that of them in few-seed scenarios. Moreover, 6Vision can function as a preliminary detection module for existing algorithms, effectively addressing the challenges they face in few-seed scenarios and guiding them toward the correct detection direction. Additionally, 6Vision achieves a remarkable conversion gain (\textit{CG}) ranging from 242\%$\sim$2,081\%.
    \item To facilitate the detection of existing algorithms in few-seed scenarios, we transform the scenarios by continuously running 6Vision, achieving a conversion rate (\textit{CR}) of 28.97\%. Additionally, we construct a dataset to enrich the IPv6 hitlist which contains 5.67M addresses. https://github.com/zwjsndy/6Vision.git 
\end{itemize}
The remainder of the paper is organized as follows. 
\S \ref{chap:background} introduces the problem definition of IPv6 target generation, the basic knowledge of IPv6, and the motivations of our work.
\S  \ref{chap:6Vision} presents the system design of 6Vision. \S  \ref{chap:evaluation} evaluates the performance of 6Vision and compares it with existing algorithms. 
\S  \ref{chap:relatework} summarizes the related work on IPv6 active address scanning.
\S  \ref{chap:conclusion} concludes the paper.

\section{background and motivations}\label{chap:background}
\subsection{Background}\label{sec:definition}
\subsubsection{Problem Definition}
The IPv6 target generation problem entails the use of a target generation algorithm, denoted as \textit{$\tau$}, which takes as input the seed addresses \textit{S} to generate the target address set \textit{C} within the budget $|C|$. The target address set \textit{C} includes active addresses \textit{T} after detection. The objective of the algorithm \textit{$\tau$} is to maximize the $HitRate$ (detection accuracy), where $C \cap T$ can be obtained by sending ICMPv6 probe packets using ZMapv6 \cite{vorhofer2020extending}:
\begin{equation}
HitRate=\frac{\left | C\cap T-C\cap S \right | }{\left | C \right | } 
\end{equation}
\subsubsection{IPv6 address structure}
An IPv6 address contains 128 bits and a unicast address consists of three parts: a global route prefix, a local subnet identifier, and an Interface Identifier (IID). An IPv6 address is typically represented by eight groups of four hexadecimal characters, each with a total of 16 bits, separated by a colon. 

\subsection{Motivations}\label{chap:few-seed}
\subsubsection{Detailed description of few-seed scenarios}
We draw inspiration from the work of AddrMiner \cite{song2022addrminer}, which employs DET for probing seed addresses of less than 10. This approach leads to a notable decrease in detection efficiency.
We first define the few-seed BGP prefixes as those with fewer than 10 seed addresses. Then, we compile all seed addresses in few-seed BGP prefixes to construct an address set. The few-seed scenarios are defined as the case where the target generation algorithm uses this address set for probing.
We amalgamated the publicly released IPv6 hitlist datasets from Gasser et al. \cite{Gasser} and Song et al. \cite{addrminer}, and identified BGP prefixes with fewer than 10 seed addresses. Subsequently, we aggregated all seed addresses from these selected BGP prefixes and employed existing efficient target generation algorithms to generate 1 million target addresses. Ultimately, we observed that the $HitRate$ of the existing efficient algorithms is less than 5\% (e.g., DET:3.73\%, 6Graph: 3.44\%, 6Tree:3.81\% ), indicating a significant decline in the accuracy of existing target generation algorithms in few-seed scenarios.

\subsubsection{Distribution of seed addresses in IPv6 hitlist}
Few-seed BGP prefixes are prevalent in the IPv6 hitlist, indicating their common occurrence within the network. However, the IPv6 hitlist fails to provide a comprehensive reflection of the current network configuration state and lacks a thorough exploration of addressing configuration patterns within few-seed BGP prefixes. This limitation is exacerbated by a severe long-tail effect, where the majority of BGP prefixes fall under the category of few-seed BGP prefixes.
Indeed, a considerable number of seed addresses are concentrated within a few BGP prefixes, resulting in a limited view of network configuration. We have compiled the IPv6 hitlist published by Gasser et al. \cite{Gasser} and Song et al. \cite{addrminer} on January 14, 2024, comprising 83.03 million seed addresses. These seed addresses are distributed across 75,053 BGP prefixes, with few-seed BGP prefixes totaling 47,773, accounting for 63.65\%.


\begin{figure*}[t!]
    \centering
    \includegraphics[width=0.85\textwidth]{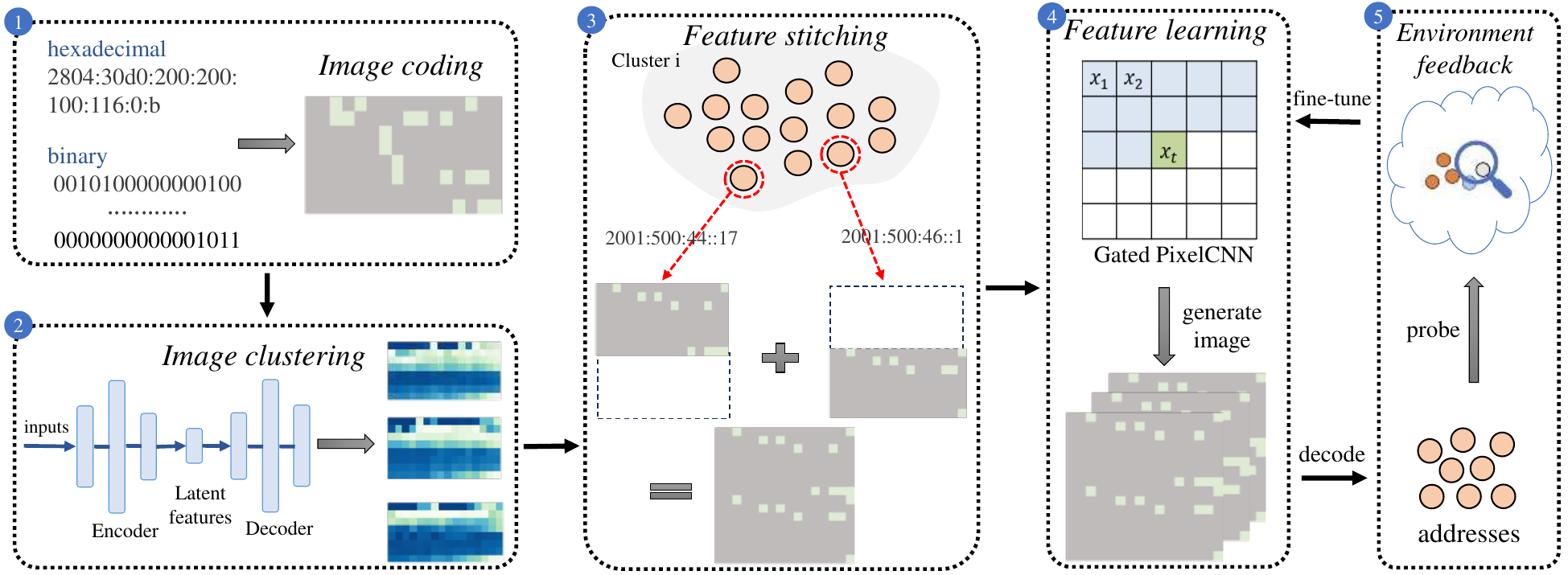}
    \caption{The overview of 6Vision and its key technical components.
}
    \label{fig:method}
\vspace{-4mm}
\end{figure*}
\section{6Vision}\label{chap:6Vision} 
\subsection{Overview of 6Vision}
The essence of our approach lies in meticulously uncovering the underlying connections of groups within an address at a more detailed level. This facilitates effective learning of address configuration rules, even when they are intricate. Furthermore, we dynamically adjust the detection direction based on feedback from active addresses, thereby helping to mitigate sampling bias.
To reveal the internal connections within an address, 6Vision encodes each address into an 8*16 image. This method combines positions with similar information within the address to uncover intricate configuration patterns. Figure \ref{fig:method} depicts the structure of 6Vision, which comprises the following five modules:

\textbf{Image encoding}:
We represent IPv6 addresses in a fine-grained binary form and rearrange them. Image encoding can place corresponding positions between groups into the same column, facilitating the model's ability to mine association relationships within different groups.

\textbf{Image clustering}:
This module clusters seed addresses with similar image features into one category. The fine-grained address division enhances the learning outcomes of subsequent models.

\textbf{Feature stitching}:
This module expands the number of learnable features and combines image encodings configured with similar addresses, thereby enabling subsequent modules to learn the relationships between addresses.

\textbf{Feature learning}:
Gated PixelCNN \cite{van2016conditional} can effectively learn image features and generate features from low to high bits. It then decodes the generated features to obtain the target addresses.

\textbf{Environment feedback}:
This mechanism can fine-tune the model based on the newly detected active addresses, continuously enhancing detection efficiency.

Initially, 6Vision converts seed addresses into binary form and encodes them as images. The encoded seed addresses are clustered into 
k classes for fine-grained learning. Secondly, 6Vision performs feature stitching in each class to combine image encodings configured with similar addresses, aiming to uncover the intricate configuration rules of seed addresses. Subsequently, 6Vision utilizes Gated PixelCNN to learn and generate image features, decoding the generated features into target addresses. Finally, 6Vision incorporates a feedback mechanism to fine-tune the model as the detection proceeds.

\subsection{Image encoding}
Our image encoding method can encode either an address, which is part of the 6Vision model structure, or a set of addresses, which is simply a visualization method.
\subsubsection{Image encoding of IPv6 address}
Existing machine-learning-based generation methods \cite{cui20216gan,cui20206gcvae,cui20216veclm} encode seed addresses into text vectors and learn the features of these text vectors. However, these methods encounter challenges in effectively mining the structural rules of address configuration. Moreover, there is an issue of ``forgetting" in long-distance text learning. Image coding can compact together locations with similar information in an address. This enables the model to better mine the configuration rules of seed addresses. In addition, each element in the image coding is a bit, which allows the model to learn address features in a finer granularity. The specific method is as follows:

As mentioned in \S \ref{sec:definition}, an IPv6 address contains eight groups. Therefore, we divide a 128-bit IPv6 address \textit{S} into 8 groups of 16 bits each:
\begin{enumerate}
    \item \textit{S} can be expressed as: $S=s_{1}s_{2}s_{3}\dots s_{128} $, where $s_{i}$ is the \textit{i}-th bit in the IPv6 address.
    \item The grouped address can be represented as $S=S_{1}S_{2}\dots S_{8} $, where each $S_{i}$ is a 16-bit substring. 
    \item The characters contained in the \textit{i}-th group $S_{i}$ can be represented as $S_{i}=s_{\left [ 16\left ( i-1 \right ) +1 \right ] } s_{\left [ 16\left ( i-1 \right ) +2 \right ] } \cdot \cdot \cdot s_{\left [ 16\left ( i \right )  \right ] } $, where $i=\left \{ 1,2,\dots 8 \right \} $.
\end{enumerate}

To enhance the mining of structural rules within address configuration, particularly the relationships between groups, we represent each group as a binary number. Each group is considered a row of the image vector, resulting in an 8-row image. This module encodes elements in the same position of different groups in a column, as elements with the same or similar positions in groups exhibit certain correlations.
The image encoding module in Figure \ref{fig:method} illustrates an example of image encoding. For an IPv6 address ``2804:30d0:200:200:100:116:0:b", it is fully written as ``280430d002000200010001160000000b", and the binary representation of the first group (``2804") is ``0010100000000100", which is depicted in the first row of the image. The binary representation of the second group is illustrated on the second row. Similarly, the image encoding of the address can be obtained. It can be observed that there are correlations between the elements in the same position of adjacent rows, such as continuous 0s or continuous 1s.

\subsubsection{Image encoding of IPv6 address set}
For an IPv6 address set, the same method can be employed for image encoding. Unlike encoding a single address, each pixel represents the entropy of that position. The entropy calculation method for each bit is as follows, where \textit{i} represents the row number (top to bottom) and \textit{j} represents the column number (left to right):
\begin{equation}
    H\left ( x_{i,j}   \right ) =-\frac{1}{4} \sum_{w\in \Omega }^{}P\left ( X_{i,j}=\omega   \right )  \cdot \log_{}{P\left ( X_{i,j}=\omega   \right )  } 
\end{equation}
\vspace{-5mm}
\begin{center}
    $\Omega=\left \{ 0,1 \right \}, i=\left \{ 1,2\cdot \cdot \cdot  8\right \}, j=\left \{ 1,2\cdot \cdot \cdot  16\right \} $
\end{center}
\vspace{-1mm}

\begin{figure}[htbp]
    \begin{subfigure}{0.21\textwidth}
        \includegraphics[width=\linewidth]{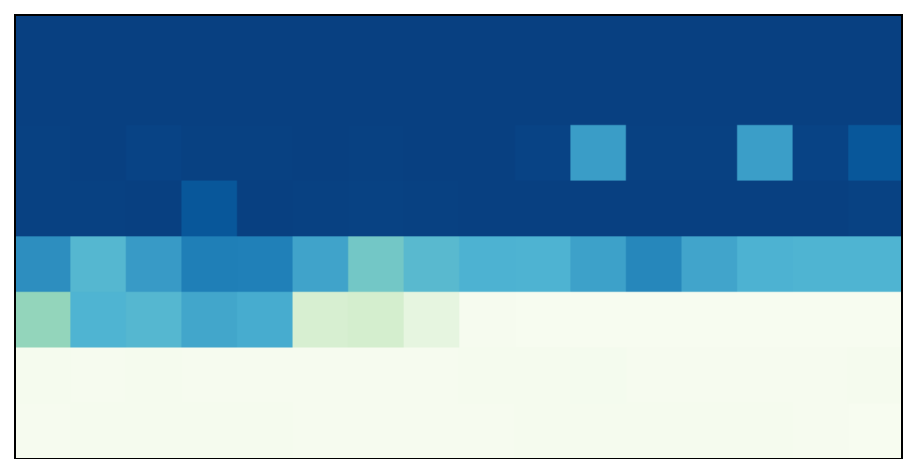}
        \caption{sufficient-seed scenarios}
        \label{fig:2a01}
    \end{subfigure}
    \hfill
    \begin{subfigure}{0.21\textwidth}
        \includegraphics[width=\linewidth]{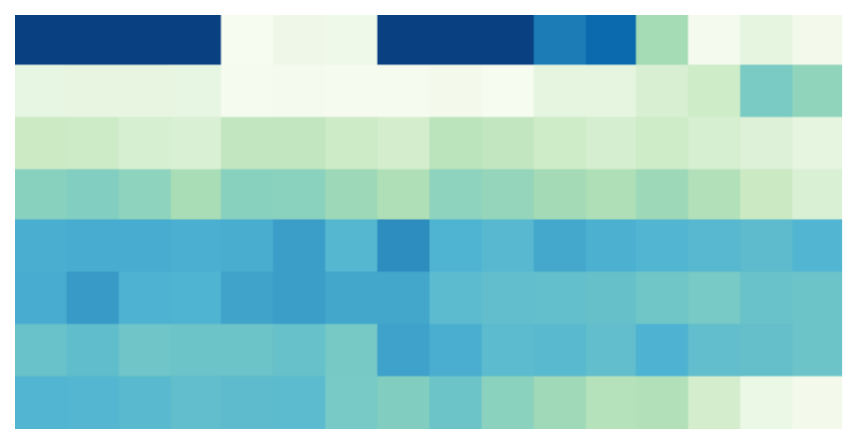}
        \caption{few-seed scenarios}
        \label{fig:few}
    \end{subfigure}
    \hfill
    \raisebox{0.02\height}{
    \begin{subfigure}{0.02\textwidth}
        \includegraphics[width=\linewidth]{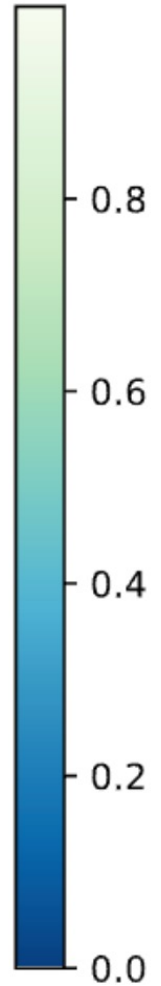}
        \label{fig:img3}
    \end{subfigure}}
    \caption{The image encoding of the IPv6 address set.}
    \label{fig:all_img}
    \vspace{-3mm}
\end{figure}

Figure \ref{fig:2a01} illustrates the image encoding of the IPv6 address set under the BGP prefixes with sufficient seed addresses, while Figure \ref{fig:few} depicts the image encoding of seed addresses in few-seed scenarios. For sufficient-seed scenarios, we select 17 BGP prefixes from the hitlist containing a large number of seed addresses and extract the same number of addresses for each BGP as in the few-seed scenarios. We repeat this process 10 times and finally calculate the average of each bit in the 170 image encoding as the final image encoding. In both figures, darker blue pixels correspond to lower entropy, while lighter green pixels correspond to higher entropy. 
The image encoding of the address set vividly illustrates that seed addresses in few-seed scenarios exhibit more intricate configuration rules, as evidenced by the higher entropy of each bit. However, BGP prefixes with a large number of seed addresses, often exhibit fixed patterns, and existing target generation algorithms are adept at discovering simple patterns.
To quantitatively illustrate the dispersion of addresses within an address set, we define the comprehensive entropy (\textit{CE}) of an address set as the sum of the entropy of all locations, normalized to the range of 0 $\sim$ 1. A higher CE indicates that the values of each bit in the address set are more dispersed, which suggests that the address configuration patterns are diverse and difficult to mine.
\begin{equation}
    CE =\frac{1}{128} \sum_{j\in J}^{}\sum_{i\in I}^{}  H\left ( x_{i,j}   \right )
\end{equation}
\begin{center}
   $ I=\left \{ 1,2\cdot \cdot \cdot  8\right \}, J=\left \{ 1,2\cdot \cdot \cdot  16\right \} $
\end{center}
\vspace{-4mm}
Quantitatively, the \textit{CE} of the two seed sets are 0.36 and 0.46, respectively. This further validates that the seed addresses exhibit intricate configuration rules in few-seed scenarios.
\subsection{Image clustering}
In few-seed scenarios, seed addresses exhibit significant differences across different prefixes, owing to the wide distribution range of addresses and intricate configuration rules. The image clustering module serves two purposes: first, by clustering images based on their addresses, the model can fine-grainedly learn the characteristics of each subclass. Second, it can group addresses with similar configuration patterns into the same cluster, facilitating the learning of the relationships between similar addresses in the subsequent feature stitching module. 
Additionally, it is essential to cluster according to the image features, considering that the addresses are encoded as images. To achieve better clustering, we use VAE \cite{kingma2013auto} to extract image features and group seed addresses with similar image features together. We opt for VAE because it can compress image features into latent features which are low-dimensional features, thereby reducing the complexity and time cost of clustering. The steps are as follows:
\begin{enumerate}
    \item Learn the image encoding of each seed address using VAE.
    \item Obtain the latent feature of each image encoding using the Encoder.
    \item Cluster the latent feature by K-means \cite{krishna1999genetic}.
\end{enumerate}

First, we learn the image encoding features of each address in the few-seed scenarios and subsequently obtain the latent feature of each image encoding. The latent feature is a low dimensional vector that contains the important information of the image \cite{kingma2013auto}. It not only greatly reduces the dimensions of the image encoding but also encapsulates important features. Therefore, we employ the K-means algorithm to cluster these latent features. Unlike directly applying the K-means algorithm to cluster the seed addresses, our clustering method fully extracts the rules of image encoding and conducts clustering based on the similarity of image encodings. This enables the subsequent generation model to learn the image features at a finer granularity. In the next three modules, we build a model for each subclass separately.

The image clustering module in Figure \ref{fig:method} illustrates 3 examples of image encoding. The \textit{CE} of these subclasses are 0.40, 0.39, and 0.33, respectively, which are lower after clustering compared to before (previously: 0.46). This suggests that our clustering method effectively narrows down the range of variations for each bit, enabling us to gather images with similar address configuration rules in a more efficient manner. Image clustering thereby facilitates the subsequent learning of image features.

\subsection{Feature stitching} 
We observe that although the configuration methods for addresses under different BGP prefixes vary, there are similarities in the addresses under some BGP prefixes. In few-seed scenarios, fully learning these patterns is challenging if we only consider addresses from a single BGP prefix. However, by clustering addresses with similar image features, we can combine the image features of addresses in each subclass for more effective learning.
Feature stitching offers two advantages. First, it expands the number of features that can be learned, thus enhancing the model's learning effectiveness. Second, it amalgamates addresses associated with configuration patterns for learning. Consequently, the introduction of feature stitching addresses the challenge of the model's inability to mine intricate configuration rules. The procedure for feature stitching is outlined as follows:

In each subclass after clustering, we randomly select 2 image encodings of addresses and concatenate them into 1 feature vector, which serves as the input feature for the subsequent model. Specifically, since an IPv6 address can be encoded into an 8*16 feature (8 rows and 16 columns), concatenating two IPv6 addresses yields a 16*16 feature (16 rows and 16 columns). For an address set with $N$ addresses, feature stitching can theoretically extend it to $2*N^2$ learnable feature vectors. The feature stitching module in Figure \ref{fig:method} illustrates an example of feature stitching using two addresses.

Feature stitching indeed increases the number of learnable features from $N$ to $2*N^2$, thereby promoting more convergent model training and enhancing the model’s learning efficiency. Additionally, the new features generated through feature stitching can capture correlations between addresses within the same subclass. Consequently, during model training, predicting two similar addresses simultaneously improves the model's capacity to learn the relationships between them.

\subsection{Feature learning}
Once the stitched features are obtained, a generative model is used to learn these feature vectors and produce new ones that follow the same distribution. These newly generated feature vectors, which encapsulate the image characteristics of target addresses, can then be decoded into IPv6 addresses that are more likely to be active. We assume that each bit in an address is determined solely by the preceding bits, aligning with the hierarchical nature of address configuration. For this reason, we utilize Gated PixelCNN to learn the image features. 

The Gated PixelCNN is a deep-learning model used for image generation. It is based on Convolutional Neural Networks (CNNs) \cite{lecun1998gradient} and incorporates a gating mechanism \cite{hochreiter1997long}, which enhances the model's expressive capabilities.
Gated PixelCNN predicts the value of each pixel in the image sequentially, proceeding from left to right and top to bottom.
Specifically, for an image \textit{x} of size $m\times n$, it can be treated as a one-dimensional sequence: $x_{1},x_{2},\cdot \cdot \cdot ,x_{m\times n} $. The probability of \textit{x} can be expressed as:
\begin{equation}
    p(x)=\prod_{i=1}^{m\times n}p(x_{i}|x_{1},x_{2},\cdot \cdot \cdot ,x_{i-1} )=\prod_{i=1}^{m\times n} p(x_{i}|x_{<i})
\end{equation}
Where \( x \) represents an image, \( x_i \) denotes the \( i^{th} \) pixel in the image, and \( x_{<i} \) represents all the pixels that come before it. This formula expresses the conditional probability distribution of each pixel in the image. Gated PixelCNN generates images by learning conditional probability distributions. To ensure that the generation of each pixel depends solely on the preceding pixels, Gated PixelCNN employs masked convolutions, which include Vertical Masks and Horizontal Masks, to extract probability features. By stacking multiple layers of masked convolutions, the receptive field of the current prediction pixel can extend to cover the entire range of pixels in the upper-left corner.

\begin{figure}[htbp]
    \centering
    \includegraphics[width=0.5\textwidth]{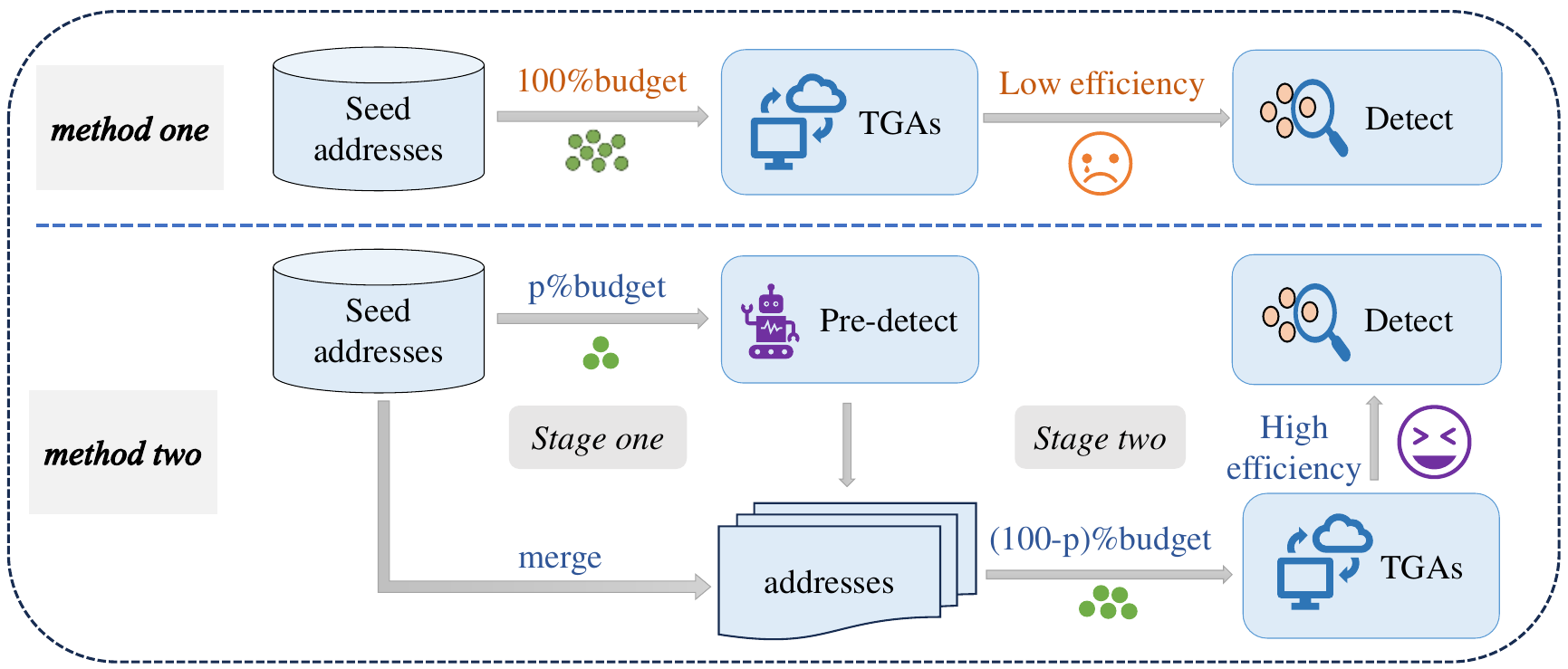}
    \caption{The definition of \textit{CG}.}
    \label{fig:two_stage}
    \vspace{-4mm}
\end{figure}
\begin{figure*}[t]
  \centering
  \begin{minipage}[t]{0.44\textwidth}
    \centering
    \includegraphics[width=\textwidth]{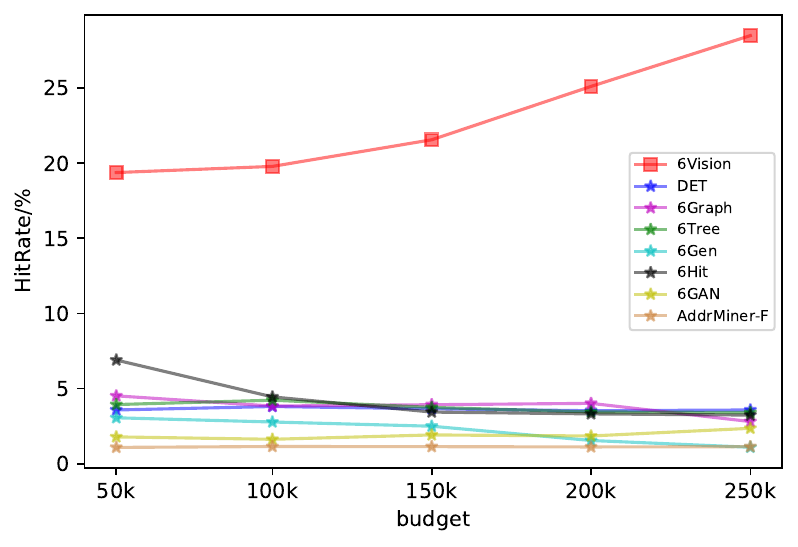}
    \caption{$HitRate$}
    \label{fig:hitrate}
  \end{minipage}
  \hfill
  \begin{minipage}[t]{0.44\textwidth}
    \centering
    \includegraphics[width=\textwidth]{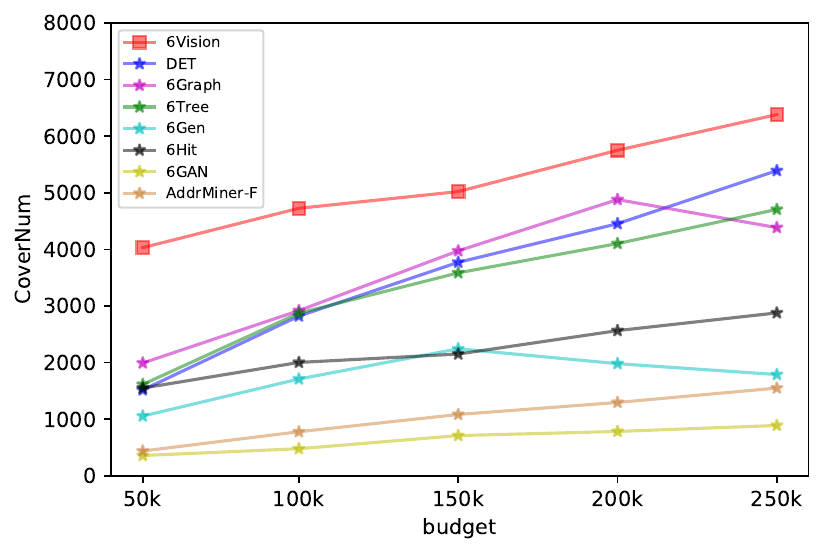}
    \caption{\textit{CoverNum}}
    \label{fig:covernum}
  \end{minipage}
  \vspace{-3mm}
\end{figure*}

\subsection{Environment feedback}
Although the model learns the configuration rules of seed addresses, it faces challenges in fully capturing the diversity of the IPv6 address space, resulting in biased sampling. Seed addresses are merely a biased sample of the real addresses. There are notable disparities between the sampled seed addresses and the true address configuration. This discrepancy is particularly pronounced in few-seed scenarios, exacerbating sampling bias. Such bias can lead the model astray, hindering its ability to accurately learn the distribution of genuine active addresses. To address this issue and enhance the efficient utilization of address detection resources, we introduce an environment feedback mechanism.
First, we train the model until convergence. During this phase, Gated PixelCNN learns the distribution of seed addresses. Subsequently, we fine-tune its parameters to better capture the characteristics of real active addresses. The fine-tuning process unfolds as follows:
In each training iteration, we leverage ZMapv6 to assess the activity of target addresses, subsequently filtering out any aliased address\cite{gasser2018clusters, cheng2024luori} and collect non-aliased active addresses. Following this, the module feeds back the image encoding of these addresses to the Gated PixelCNN for a brief retraining.
Through multiple feedback iterations, we aim to alleviate the sampling bias inherent in seed addresses and steer the model towards generating target addresses more likely to be active. The incorporation of an environment feedback mechanism enables the model to better understand the distribution of real active addresses, thereby enhancing the accuracy of the generated target addresses.

\section{evaluation}\label{chap:evaluation}
\subsection{Experimental setup}

\subsubsection{Evaluation index setting}
We formulate the following evaluation metrics: \textit{HitRate} and \textit{CoverNum} are employed to assess the accuracy and coverage of the algorithm, while the Conversion Rate (\textit{CR}) and Conversion Gain (\textit{CG}) gauge the capability to convert few-seed scenarios and the enhancement in the performance of existing algorithms post-conversion, respectively.

\begin{itemize}
    \item \textit{HitRate} metric assesses the accuracy of the target generation algorithm. We have already defined \textit{HitRate} in \S~\ref{sec:definition}.
    \item \textit{CoverNum} metric represents the number of BGP prefixes covered by the active target addresses.
    \item The \textit{CG} (Conversion Gain) metric evaluates the efficiency improvement of existing target generation algorithms when detecting in few-seed scenarios. As shown in Figure \ref{fig:two_stage}, it measures the increase in \textit{HitRate} achieved by employing a two-stage detection approach compared to using existing target generation algorithms alone. In the two-stage approach, a preliminary detection module $pre$ (e.g., 6Vision) is first utilized to consume a percentage $p\%$ of the budget for the initial detection phase, enabling the conversion of few-seed scenarios. Subsequently, the remaining budget $(100-p)\%$ is allocated for the second stage detection using target generation algorithms $\tau$ (Method two). The active addresses identified by the preliminary detection module, along with the initial seed addresses, serve as input for the second stage. The enhancement in \textit{HitRate} observed with the two-stage detection approach, in contrast to using existing target generation algorithms $\tau$ alone (Method one), defines the Conversion gain (\textit{CG}), where $HitRate_{j}^i$ represent the detection accuracy use algorithm $j$  and method $i$: 
    \begin{equation}
        \small
CG=\frac{p\%*HitRate_{pre}^2+(100-p)\%*HitRate_{\tau}^2 }{HitRate_{\tau}^1}-1
    \end{equation} 
     \item \textit{CR} evaluates the conversion ratio of few-seed BGP prefixes, where \textit{B} represents the BGP set, and \textit{p} represents each BGP prefix in \textit{B}. The indicator function $I(x)$ is utilized to count the number of conditions met; when \textit{x} is true, it yields a value of 1, otherwise, it yields 0.
    \begin{equation}
        CR=\frac{\sum_{p\in B}^{} I(\left | p \right | >10)}{\left | B\right | } 
    \end{equation}
\end{itemize}
\begin{figure*}[htbp]
    \centering
    \begin{subfigure}{0.47\textwidth}
        \includegraphics[width=\linewidth]{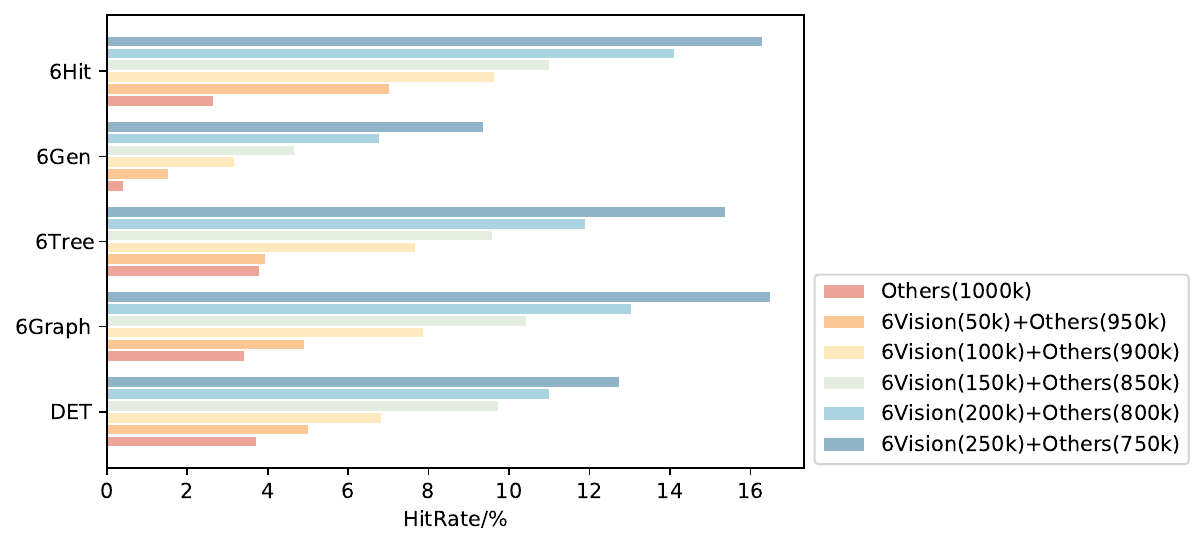}
        \caption{\textit{CG} of 6Vision}
        \label{fig:conver1}
    \end{subfigure}
    \hfill
    \begin{subfigure}{0.47\textwidth}
        \includegraphics[width=\linewidth]{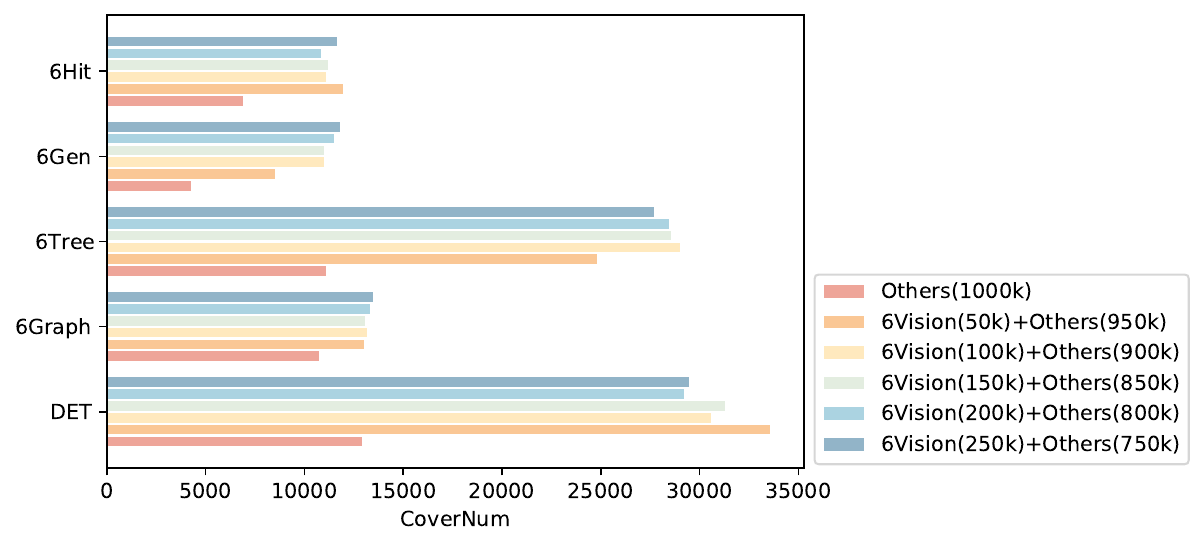}
        \caption{\textit{CoverNum} increase of 6Vision}
        \label{fig:conver2}
    \end{subfigure}
    \hfill
     \caption{The \textit{CG} of 6Vision under different values of \textit{p}.}
     \vspace{-5mm}
\end{figure*}

\subsubsection{Active address verification method}
 We utilize the ZMapv6 tool \cite{Zmapv6} for scanning purposes. Prior research has demonstrated that active addresses exhibit the highest response probability to the ICMPv6 protocol \cite{song2022det}. Hence, we opt for the ICMPv6 protocol for our detection methodology. A single packet is dispatched to each address; upon receiving a response, the address is classified as active. Furthermore, we employ APD \cite{gasser2018clusters} to eliminate aliased addresses from the dataset.
\subsubsection{Parameter selection}
We aggregated the seed addresses from the few-seed BGP prefixes discussed in \S \ref{chap:few-seed} as input, which collectively totals 149,795 seed addresses. We classified the image encoding of seed addresses into 6 subcategories, each subcategory uses VAE to extract image features, and we train VAE for 200 epochs. Subsequently, the clustered image encodings were arranged in dictionary order. Each image encoding was then concatenated with the following five images in sequence (selected based on the smallest difference in dictionary order). During the training of the Gated PixelCNN model, we utilized a 3*3 convolutional kernel and set the training batch size to 64. The training process spanned 40 epochs to ensure comprehensive learning. To expedite the generation of image encoding, we generated images with dimensions of 8*16, which can be effectively decoded into an address.
In the environment feedback mechanism, feedback is initiated after generating 25k target addresses, followed by 10 epochs of fine-tuning. 
\subsubsection{Ethical considerations}
We follow ethical conventions for network measurement,
including recommendations provided by Partridge et al. \cite{partridge2016ethical} and Dittrich et al. \cite{kenneally2012menlo}.We limited the packet-sending rate to 10 Mbps, which is unlikely
to have an impact on the performance of the network.
\subsection{Detection performance evaluation}\label{sec:hitrate}
We compare six efficient target generation algorithms (DET\cite{song2022det}, 6Graph\cite{yang20226graph}, 6Tree\cite{liu20196tree}, 6Gen\cite{murdock2017target}, 6Hit\cite{hou20216hit}, 6GAN\cite{cui20216gan}, Addrminer-F\cite{song2022addrminer}) with 6Vision. 6Scan\cite{hou20236scan} operates on the same principle as 6Tree for small-scale detection. 6Forest\cite{yang20226forest} identifies many seed addresses as outliers, leading to a significant reduction in detection accuracy. Therefore, we did not choose these two algorithms. Figure \ref{fig:hitrate} presents the comparison results. The \textit{HitRate} of 6Vision is improved by 181\% $\sim$ 2,490\% compared to existing algorithms. With increasing budgets, the $HitRate$ of 6Vision progressively rises. Upon reaching a budget of 250k, the $HitRate$ attains 28.48\%, showcasing superior detection efficacy compared to other algorithms in few-seed scenarios. Notably, it achieves a staggering 696\% increase over the second-place algorithm (DET, $HitRate$: 3.58\%).
The substantial enhancement in the $HitRate$ of 6Vision can be attributed to its key design elements. Firstly, the image encoding method allows for intricate analysis of intricate configuration rules. Secondly, feature stitching addresses the challenge of low learning efficiency stemming from the intricate configuration rules in few-seed scenarios. Lastly, the environment feedback mechanism dynamically adjusts model parameters during detection, thereby alleviating sampling bias associated with seed addresses.

The higher $HitRate$ achieved by 6Hit compared to other algorithms at budgets of 50k and 100k can be attributed to its innovative mechanisms, namely ``nodes chipping" and ``space repartition" \cite{hou20216hit}. These mechanisms effectively address sampling bias during the initial detection stage and allow 6Hit to dynamically adjust the target region generation based on the expected reward of each region. However, as the budget increases to 250k, the $HitRate$ of 6Graph experiences a sharp decline. This decline may be attributed to the limitations of the graph-based mining method, which can result in erroneous clustering of seed addresses and subsequently lead to inaccurate pattern generation. The limitations of 6GAN stem from its approach of encoding seed addresses as texts for training, which poses challenges in effectively learning the configuration patterns of these addresses. Consequently, the target addresses generated by 6GAN exhibit a high degree of randomness, leading to inefficiency in the detection process. Similarly, Addrminer-F faces difficulties in building a comprehensive pattern library due to the limited number of seed addresses available in few-seed scenarios. As a result, Addrminer-F experiences a significant decline in efficiency, highlighting the importance of addressing the challenges posed by few-seed scenarios in target address generation algorithms.

Using \textit{CoverNum} as an evaluation metric allows us to assess the breadth of coverage achieved by the explored active addresses. A higher \textit{CoverNum} value indicates that the addresses span across a greater number of BGP prefixes, indicative of a more exploratory approach. As depicted in Figure \ref{fig:covernum}, the number of BGP prefixes covered by 6Vision steadily increases with the augmentation of the budget. The \textit{CoverNum} is 1.18$\sim$11.20 times that of the existing algorithms. This suggests that 6Vision not only uncovers more active addresses but also spans across a larger number of BGP prefixes, rather than being concentrated on just a few. Thus, 6Vision maintains high accuracy while showcasing remarkable exploratory capabilities.
6Graph and 6Gen exhibit an inflection point as the budget increases, indicating their exploration tends to concentrate more on high-density areas, i.e., they are constrained to a small number of BGP prefixes as the budget increases. On the other hand, 6GAN and Addrminer-F yield lower \textit{CoverNum} because they uncover fewer active addresses. The high \textit{CoverNum} achieved by 6Vision can be attributed to its proficiency in mining the configuration patterns of intricate addresses. Additionally, the incorporation of feature stitching enables the model to grasp the correlations between addresses, consequently enhancing its exploratory capacity.
\subsection{Conversion ability evaluation}
The \textit{CG} metric allows us to assess the enhanced performance of existing algorithms due to the incorporation of 6Vision. We tested the \textit{p} values of 5, 10, 15, 20, and 25, with a budget of 1 million. Figure \ref{fig:conver1} illustrates that when \textit{p} is set to 25, the \textit{CG} of 6Vision ranges from 242\% $\sim$ 2,081\%. This signifies a substantial improvement in the detection accuracy of other algorithms.
Among them, the performance of 6Gen has been significantly improved. Combining 6Vision with 6Gen for detection resulted in a notable increase in the overall $HitRate$, soaring from 0.43\% to 9.38\%. This underscores the efficacy of employing 6Vision as a preliminary detection module, which not only enriches the addresses in few-seed scenarios but also consumes only a fraction of the budget. By mitigating the sampling bias of seed addresses and enriching addresses with intricate configuration rules, 6Vision significantly enhances the learning process of existing algorithms.
Figure \ref{fig:conver2} highlights an interesting observation: while the \textit{HitRate} of the two-stage method displays noticeable improvements, the \textit{CoverNum} exhibits a different trend. Unlike \textit{HitRate}, which gradually increases with varying values of \textit{p}, the \textit{CoverNum} experiences a substantial surge followed by fluctuations around a specific value. This pattern indicates that 6Vision has the capability to significantly enhance the coverage of other algorithms even with a limited budget.
Indeed, the \textit{CoverNum} of each target generation algorithm reflects its performance characteristics. DET, known for its efficient detection capabilities and broad coverage, exhibits a notably higher \textit{CoverNum} (33617, when \textit{p} is set to 25) compared to other algorithms.
\begin{figure*}[htbp]
    \begin{subfigure}{0.49\textwidth}
        \includegraphics[width=\linewidth]{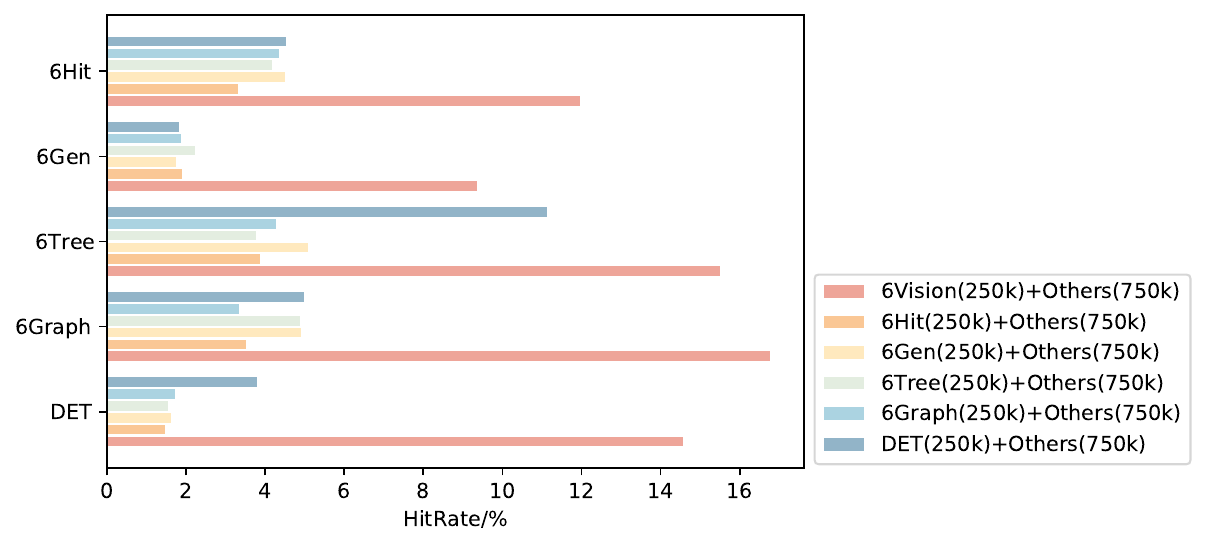}
        \caption{Comparison of \textit{CG}}
        \label{fig:conbgp1}
    \end{subfigure}
    \hfill
    \begin{subfigure}{0.49\textwidth}
        \includegraphics[width=\linewidth]{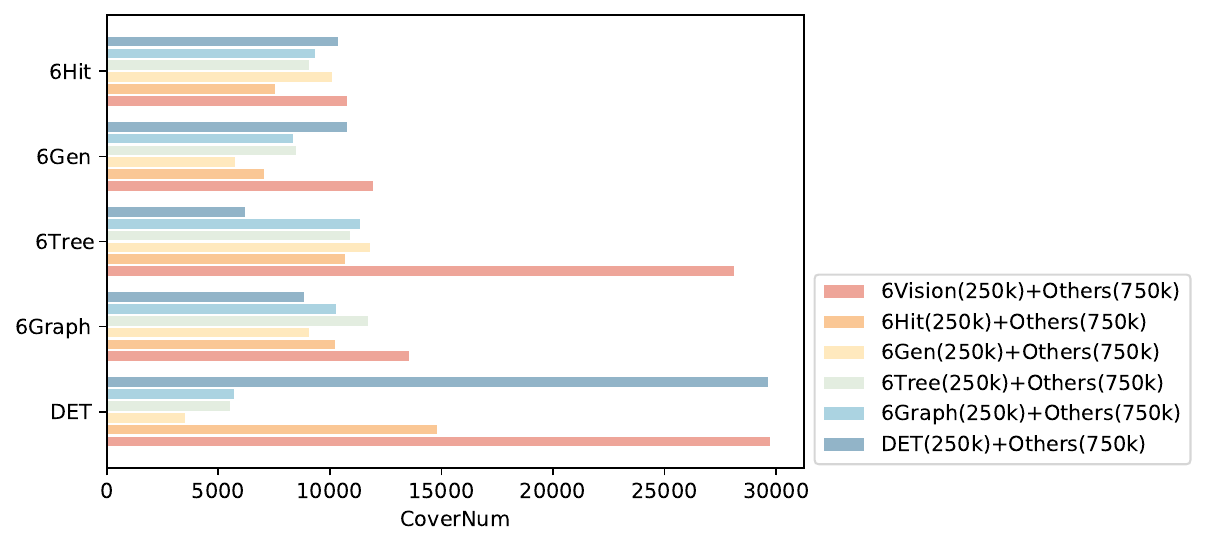}
        \caption{Comparison of \textit{CoverNum}}
        \label{fig:conbgp2}
    \end{subfigure}
    \caption{The \textit{CG} of different algorithms (values of \textit{p} is 0.25).}
    \label{fig:conbgp}
     \vspace{-3mm}
\end{figure*}
Using both \textit{CG} and \textit{CoverNum} as comparative metrics provides a comprehensive evaluation of 6Vision's impact on transforming few-seed scenarios. While \textit{CG} highlights the enhancement in detection accuracy achieved by integrating 6Vision with existing algorithms, \textit{CoverNum} emphasizes the expanded coverage of BGP prefixes, showcasing the broader exploration capabilities facilitated by 6Vision.
Employing 6Vision and five other high-accuracy algorithms as preliminary detection modules, we evaluate their effectiveness in transforming few-seed scenarios. Initially, each preliminary detection module is allocated 25\% of the total budget to generate addresses. Subsequently, the existing algorithms leverage the newly discovered addresses in conjunction with the seed addresses for detection, utilizing the remaining budget of 0.75 million. This approach allows for a comprehensive assessment of 6Vision's impact on enhancing detection accuracy and broadening BGP prefix coverage in few-seed scenarios.

Figure \ref{fig:conbgp1} illustrates the comparison of the Conversion Gain (\textit{CG}) achieved by 6Vision and other algorithms when used as preliminary detection modules for existing algorithms. Both approaches contribute to enhancing the detection accuracy of the existing algorithms, yet 6Vision exhibits a significantly superior performance, surpassing other algorithms by 1.39 to 9.79 times. Figure \ref{fig:conbgp2} compares the \textit{CoverNum} of different algorithms. The results clearly demonstrate that 6Vision substantially enhances the coverage of BGP prefixes for all algorithms. By achieving a high $HitRate$ and distributing detected active addresses across a larger number of BGP prefixes, 6Vision significantly amplifies the effectiveness of existing algorithms when employed as a preliminary detection module. In contrast, other algorithms struggle to effectively handle few-seed scenarios. While DET, when used as a preliminary detection module, outperforms other algorithms, it heavily relies on the input addresses. However, when DET serves as the detection model in the second stage, the detection accuracy and coverage are significantly reduced, except when 6Vision is employed as the preliminary detection model.

\begin{table}[tbp]
  \caption{The \textit{HitRate} in the ablation experiment.}
  
  \begin{center}
    \begin{tabular}{|c||c|c|c|c|}
      \hline
      \textbf{Budget} & \textbf{w/ f$^{\mathrm{a}}$ w/o s$^{\mathrm{b}}$} & \textbf{w/o f w/ s} & \textbf{w/o f w/o s} & \textbf{w/ f w/ s} \\
      \hline\hline
      50k & 6.39\% & 5.95\% & 5.09\% & 19.37\% \\
      \hline
      100k & 7.91\% & 5.42\% & 4.88\% & 19.78\% \\
      \hline
      150k & 10.81\% & 5.12\% & 4.69\% & 21.55\% \\
      \hline
      200k & 9.69\% & 5.07\% & 3.88\% & 25.10\% \\
      \hline
      250k & 9.39\% & 5.05\% & 3.76\% & 28.48\% \\
      \hline
      \multicolumn{5}{l}{$^{\mathrm{a}}$environment feedback module, $^{\mathrm{b}}$feature stitching module}
    \end{tabular}
  \end{center}
  \label{tab:ablation}
  \vspace{-8mm}
\end{table}

\subsection{Ablation experiment}
To better elucidate the significance of modules in 6Vision, we conducted ablation experiments to assess their impact on detection efficiency. As shown in TABLE \ref{tab:ablation}, we observed a decrease in the $HitRate$ of 6Vision as the budget increased when the environmental feedback mechanism was excluded. This suggests that the environmental feedback mechanism plays a crucial role in mitigating the sampling bias of seed addresses. By continuously fine-tuning the model parameters as detection progresses, 6Vision consistently discovers active addresses. 
Excluding feature stitching from the model resulted in a significant decrease in the $HitRate$ when the budget exceeded 150k. This decline can be attributed to the model's inability to effectively learn image features, especially when fine-tuning with newly discovered addresses. Additionally, without feature stitching, the model fails to capture associative relationships between similar patterns of addresses, leading to a diminished ability to uncover new active addresses.
The inclusion of the feedback mechanism without feature stitching resulted in a fluctuating trend in the $HitRate$ as the budget increased. Initially, the feedback mechanism notably enhanced the model's accuracy, particularly when the budget was small. However, as the budget expanded, the model's learning effectiveness progressively declined, ultimately causing a decrease in the $HitRate$.

In summary, these two modules complement each other effectively. Feature stitching significantly enhances the model's learning capability and exploratory ability, enabling it to uncover the \textbf{intricate configuration rules} of seed addresses. On the other hand, the environmental feedback mechanism improves the model's ability to sustain efficient detection by alleviating the \textbf{sampling bias} of seed addresses.

\subsection{Case study of 6Vision}
\begin{figure}[t!]
    \centering
    \includegraphics[width=0.48\textwidth]{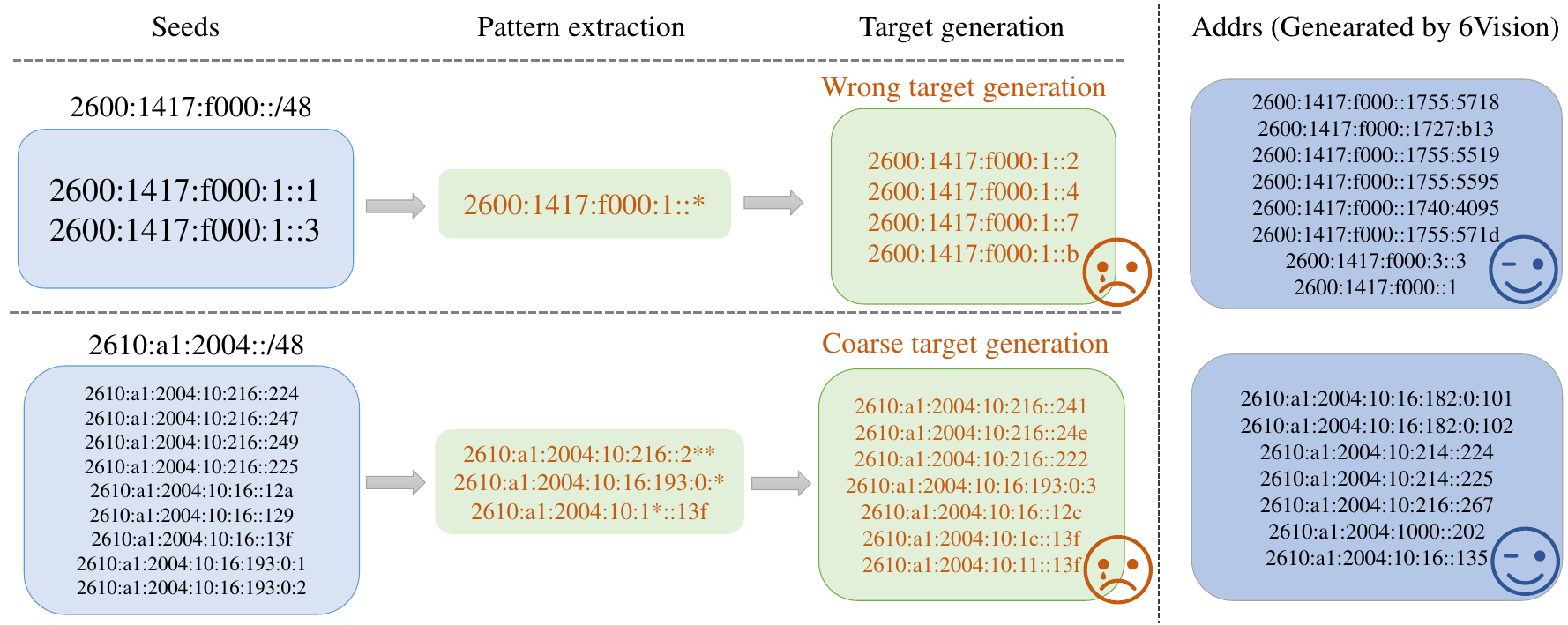}
    \caption{
    Case study of 6Vision.}
    \label{fig:problem}
    \vspace{-4mm}
\end{figure}

To better understand the reasons for 6Vision's high hit rate, let's take two examples to compare 6Vision with existing target generation algorithms.
As shown in Figure \ref{fig:problem}, there are two problems in the few-seed scenarios. 

\textbf{Wrong target generation}: 
Severe sampling bias of seed addresses is evident, whereby the distribution of seed addresses differs significantly from that of active addresses in the actual network. This bias often misleads pattern extraction. For instance, consider the prefix ``2600:1417:f000::/48", which only has two seed addresses. Existing target generation algorithms might erroneously learn the pattern ``2600:1417:f000:1::*", resulting in the generation of target addresses that are not active.

\textbf{Coarse target generation}: The intricate configuration rules present a challenge for existing methods to extract precise patterns. Consider the prefix ``2610:a1:2004::/48", which contains only 9 seed addresses but features 7 variable nybbles. This complexity makes it challenging for existing target generation algorithms to cluster these addresses into high-density regions. Consequently, numerous patterns are extracted, leading to the generation of target addresses that are not active when utilizing these patterns.

It is noteworthy that the addresses generated by 6Vision exhibit significant differences from the initial addresses, owing to its two core designs. The first design is the feature stitching module, which enables the model to learn associations between addresses under similar BGP prefixes. For instance, in the generation of the address ``2600:1417:f000::1740:4095", the model learns the configuration of ``1740" from addresses such as ``2600:1417:3800:2::1740:8996" and ``2600:1417:3800:3::1740:89c7". The second design is the environmental feedback module, which enables the model to mitigate sampling bias and adjust the probe direction based on the distribution of real active addresses. This adjustment is less affected by the initial seed address. In contrast, existing algorithms struggle to associate addresses under similar BGP prefixes due to significant nybble differences between different BGP addresses, making pattern extraction challenging. Additionally, patterns extracted by existing algorithms are highly influenced by the seed addresses.
\subsection{Dataset construction}

\begin{table}[h]
  \centering
    \caption{Dataset statistics}
  \resizebox{0.42\textwidth}{!}{ 
    \begin{tabular}{|c||c|c|c|}
      \hline
      \textbf{Addr} & \textbf{BGP prefixes} & \textbf{ASes} & \textbf{CR} \\
      \hline \hline
      5.67M & 47,773 & 17,433 & 28.97\% \\
      \hline
    \end{tabular}
  }
  \label{tab:IPv6 hitlist Patch}
\end{table}

To enable existing target generation algorithms to conduct large-scale detection in few-seed scenarios, we undertook the transformation of these scenarios through extended runs of 6Vision. Eventually, we constructed a dataset to enrich the IPv6 hitlist, which augmented the addresses within the existing IPv6 hitlist in few-seed scenarios and converted 13,842 few-seed BGP prefixes, achieving a conversion rate of 28.97\%. The statistics of our dataset are detailed in TABLE \ref{tab:IPv6 hitlist Patch}. To optimize the use of detection resources, we removed all aliased addresses. However, for BGP prefixes solely comprising aliased addresses, we retained a subset of 10 addresses. This strategy prevents wastage of detection resources on aliased prefixes and preserves a portion of address information under these prefixes for future use. The dataset we constructed transforms 28.97\% of the currently discovered 47,773 few-seed BGP prefixes. The richness of seed addresses under few-seed BGP prefixes is more meaningful, as they often exhibit more and harder-to-mine patterns. The discovery of these addresses is of greater significance for IPv6 network measurement and security analysis.

\section{Related work}\label{chap:relatework}
Previous research in IPv6 active search has predominantly concentrated on designing target generation algorithms that leverage seed addresses to produce target addresses. These methods operate under the assumption that seed addresses provide valuable insights into the addressing schemes employed, implying that seed data plays a pivotal role in revealing additional addresses.
The research on IPv6 target generation algorithms can be categorized into two primary approaches: statistical generation methods  \cite{ullrich2015reconnaissance,foremski2016entropy,murdock2017target,liu20196tree,hou20216hit,hou20236scan,gasser2018clusters,song2022det,yang20226graph,yang20226forest,song2022addrminer} and machine-learning-based generation methods \cite{cui20206gcvae,cui20216veclm,cui20216gan}.
\subsection{Statistical generation methods}
Statistical generation methods can be categorized into density-based generation methods \cite{ullrich2015reconnaissance,murdock2017target,song2022addrminer,song2020towards,song2024fast}, structure-based generation methods \cite{foremski2016entropy}, \cite{gasser2018clusters} and hierarchy-based generation methods \cite{liu20196tree,hou20236scan,hou20216hit,yang20226graph,yang20226forest}.

Density-based generation methods refer to generating target addresses by learning the density characteristics of seed addresses.
Ullrich et al. \cite{ullrich2015reconnaissance} proposed an algorithm for automatic pattern generation, marking the advent of this approach.
6Gen \cite{murdock2017target} uses a bottom-up clustering method to group seed addresses into high-density regions, facilitating pattern extraction and fine-grained address generation. 
DET \cite{song2022det} utilizes entropy as the segmentation index, employing a top-down approach to partition the density space tree using the minimum entropy method.
Building upon this foundation, Addrminer-S \cite{song2022addrminer} integrates a dynamic feedback strategy employing Thompson sampling \cite{russo2018tutorial} to enhance efficiency further.

Structure-based generation methods refer to the analysis of the internal structural relationships and configuration features of seed addresses, then generating addresses based on statistical relationships between their internal structures. Entropy/IP \cite{foremski2016entropy}, proposed by Foremski, uses information theory and machine learning methods to model IPv6 addresses probabilistically. 


Hierarchy-based generation methods categorize seed addresses into distinct subnet spaces based on their hierarchical characteristics. 6Tree \cite{liu20196tree} regards IPv6 addresses as high-dimensional vectors and conducts top-down hierarchical clustering (DHC) on seed addresses to construct hierarchical space trees in linear time. 6Hit \cite{hou20216hit} adds a reinforcement learning mechanism into the hierarchical space tree to allocate exploration resources according to the scanning reward for each area. 6Scan \cite{hou20236scan} optimized the detection speed of 6Tree. This method leverages the regional identifier encoding to quickly adjust the search direction without excessive computation. 6Graph \cite{yang20226graph} uses a graph theory pattern mining algorithm to find address patterns and filter out ``outliers" . 6Forest \cite{yang20226forest} uses an enhanced isolated forest algorithm to automatically filter out outlier seed addresses, mitigating the effect of ``outliers".
\subsection{Machine-learning-based generation methods}
The core idea of these methods is to treat the seed addresses as text and transform the target address generation problem into a text generation problem. They use the generation model in machine learning to learn the features of the text (seed addresses) and generate new text (target addresses). 6VecLM \cite{cui20216veclm} map IPv6 addresses to a vector space to mine the semantic relationships and uses Transformer \cite{vaswani2017attention} to build an IPv6 language model and generate target addresses. 6GCVAE \cite{cui20206gcvae} learns the address structure by stacking the gated convolutional layer \cite{dauphin2017language} to construct Variational Autoencoder (VAE) \cite{kingma2013auto}. 6GAN initially employs entropy clustering \cite{gasser2018clusters} to perform a fine-grained classification of seed addresses and then constructs target generation algorithms based on generative adversarial net (GAN) \cite{goodfellow2020generative} and reinforcement learning. 
However, encoding seed addresses as text vectors fails to effectively capture the structure of address configuration as it overlooks the potential correlation between similar positions within the internal information of each seed.

In conclusion, statistical generation methods and machine-learning-based generation methods have the following problems when detecting in few-seed scenarios. 
1) Due to the intricate configuration rules of seed addresses, it is difficult for these methods to learn the intricate configuration rules of seed addresses effectively. 
2) The sampling bias significantly influences their efficiency. Especially in few-seed scenarios, the sampling bias of seed addresses is intensified, further diminishing their detection efficiency.

\section{Conclusion}\label{chap:conclusion}

In this paper, we introduce 6Vision, a novel approach that encodes IPv6 addresses as images for the first time. 6Vision amalgamates addresses with similar configuration patterns through the feature stitching module, facilitating a precise understanding of intricate configuration rules. Additionally, 6Vision integrates an environmental feedback mechanism to counteract the effects of sampling bias, enabling efficient target generation even in scenarios with limited seed data. 6Vision demonstrates a substantial improvement in accuracy ranging from 181\%$\sim$2,490\%, along with a notable increase in \textit{CoverNum} by 1.18$\sim$11.20 times that of the existing algorithms. Moreover, 6Vision serves as an effective preliminary detection module for existing algorithms, yielding a conversion gain (\textit{CG}) of 242\%$\sim$2,081\%. Ultimately, our approach facilitates the transformation of few-seed scenarios, achieving a conversion rate (\textit{CR}) of 28.97\% and establishing the dataset to enrich the IPv6 hitlist. These advancements are promising for IPv6 network measurement and security analysis, providing better support for key challenges in the IPv6 landscape.

\section*{Acknowledgment}
We thank the anonymous ICNP reviewers and our shepherd Yashar Ganjali for their feedback and suggestions. This work is supported by the National Key R\&D Program of China under Grant 2023YFB3107202, Quan Cheng Laboratory (Grant No. QCLZD202304-2), the Research Project of Provincial Laboratory of Shandong, China(Grant No. SYS202201), the National Natural Science Foundation of China under Grant No. 62302253 and 62172251 and the Beijing Natural Science Foundation under Grant No. 4222026.
Jiahai Yang and Guanglei Song are the corresponding authors.

\bibliography{ref.bib}

\bibliographystyle{IEEEtran}

\end{document}